%% file: bare_jrnl.tex
\documentclass[journal]{IEEEtran}
\usepackage{cite}
\usepackage{url,hyperref,booktabs}
\usepackage{pifont}

\ifCLASSINFOpdf
  \usepackage[pdftex]{graphicx}
  \DeclareGraphicsExtensions{.pdf,.jpeg,.png}
  \usepackage{epstopdf}
  \usepackage{subcaption}
\else
  \usepackage[dvips]{graphicx}
\fi
\usepackage{amsmath}
\usepackage{amssymb}
\usepackage{multirow}
\usepackage[switch]{lineno}
\usepackage{array}

\usepackage{stfloats}
\usepackage[symbol]{footmisc}

\makeatletter
\def\endthebibliography{%
  \def\@noitemerr{\@latex@warning{Empty `thebibliography' environment}}%
  \endlist
}
\makeatother

\begin{document}
%
\title{Monaural Speech Enhancement Using a Multi-Branch
Temporal Convolutional Network
}
%
%
%


\author{Qiquan~Zhang,~
        Aaron Nicolson,~
        Mingjiang~Wang,~
        Kuldip K. Paliwal,~
        and Chenxu Wang
\thanks{Manuscript received date; revised date. This work was supported by the Basic Research Discipline Layout Project of Shenzhen under Grant JCYJ20170412151226061 and Grant JCYJ20170808110410773. (Joint corresponding authors: Aaron Nicolson, Mingjiang Wang.)} 
\thanks{Qiquan Zhang and Mingjiang Wang are with the School of Electronic \& Information Engineering, Harbin Institute of Technology, Shenzhen 518055, China (email: zhangqiquan\_hit@163.com; mjwang@hit.edu.cn).}
\thanks{Aaron Nicolson and Kuldip K. Paliwal are with the Signal Processing Laboratory, Griffith University, Brisbane, Queensland 4111, Australia (email: aaron.nicolson@griffithuni.edu.au; k.paliwal@griffith.edu.au).}
\thanks{Chenxu Wang is with the School of Information Science \& Engineering, Harbin Institute of Technology, Weihai 264209, China (email: wangchenxu@hit.edu.cn).}}

%
%

\markboth{Journal of \LaTeX\ Class Files,~Vol.~14, No.~8, August~2019}%
{Shell \MakeLowercase{\textit{et al.}}: Bare Demo of IEEEtran.cls for IEEE Journals}
%



\maketitle

\begin{abstract}
Deep learning has achieved substantial improvement on single-channel speech enhancement tasks. However, the performance of multi-layer perceptions (MLPs)-based methods is limited by the ability to capture the long-term effective history information. The recurrent neural networks (RNNs), e.g., long short-term memory (LSTM) model, are able to capture the long-term temporal dependencies, but come with the issues of the high latency and the complexity of training.
To address these issues, the temporal convolutional network (TCN) was proposed to replace the RNNs in various sequence modeling tasks. In this paper we propose a novel TCN model that employs multi-branch structure, called multi-branch TCN (MB-TCN), for monaural speech enhancement. 
The MB-TCN exploits split-transform-aggregate design, which is expected to obtain strong representational power at a low computational complexity. 
Inspired by the TCN, the MB-TCN model incorporates one dimensional causal dilated CNN and residual learning to expand receptive fields for capturing long-term temporal contextual information. 
Our extensive experimental investigation suggests that the MB-TCNs outperform the residual long short-term memory networks (ResLSTMs), temporal convolutional networks (TCNs), and the CNN networks that employ dense aggregations in terms of speech intelligibility and quality, while providing superior parameter efficiency.
Furthermore, our experimental results demonstrate that our proposed MB-TCN model is able to outperform multiple state-of-the-art deep learning-based speech enhancement methods in terms of five widely used objective metrics.


\end{abstract}


%
\IEEEpeerreviewmaketitle

\section{Introduction}
%
%
%
%

\IEEEPARstart{S}{peech} signals are inevitably degraded by background noise. The objective of speech enhancement is to remove the background interference and improve the overall perceived quality and intelligibility of the degraded speech. Speech enhancement is an important and challenging task in the speech processing community and is applied to a wide range of speech-related applications, for example, robust automatic speech recognition, speaker identification, mobile speech communication, and hearing aids. In this study, we focus on monaural (single-channel) speech enhancement \cite{hendriks2013dft, P.C.Loizou2013, wang2018supervised}. 

Conventional monaural speech enhancement methods are designed based on several assumptions about the speech and noise characteristics, including spectral subtraction algorithms \cite{boll1979suppression, Loizou2002multi}, Wiener filtering \cite{wiener1996, 1979enhancement}, statistical model-based methods \cite{Y.Ephraim1984, Y.Ephraim1985, Kavalekalam2019,zhang2019novel}, and non-negative matrix factorization methods \cite{NMF2009nonnegative, NMF2013supervised}. Recently, deep neural network (DNN)-based monaural speech enhancement methods \cite{wang2018supervised} have received a tremendous amount of attention since they have demonstrated a significant performance improvement over traditional approaches. 

Inspired by the concept of T-F masking in computational auditory scene analysis (CASA) \cite{CASA2006wang}, multi-layer perceptrons (MLPs) are firstly introduced by Wang and Wang \cite{wang2013towards} to estimate the ideal binary mask (IBM) \cite{IBMwang2005ideal}. Subsequently, researchers have proved that ideal ratio mask (IRM) \cite{IRMwang2014training} based methods are able to attain better speech quality than binary mask based methods. More recently, Williamson \textit{et al.} designed the complex ideal ratio mask (cIRM) \cite{cIRM} to recover both the amplitude and phase spectral of clean speech. Different from the masking-based methods, for mapping-based methods the DNN is trained to estimate the spectral features of clean speech from that of noisy speech. In \cite{xu2013experimental} Xu \textit{et al.} proposed to use an MLP to map the noisy speech log-power spectra (LPS) to the clean speech LPS. Han \textit{et al.} \cite{han2015learning} trained an MLP to learn a spectral mapping from the magnitude spectrum of noisy speech to that of clean speech. 

Recently, multiple deep learning methods have been proposed to improve the performance of statistical model-based methods \cite{xia2018priori,rehr2017importance,NICOLSON201944,zhangDeep}. The statistical model-based speech enhancement methods heavily depend on the estimate of the a \textit{priori} SNR. More recently, a residual LSTM (ResLSTM) network was proposed to estimate the a \textit{priori} SNR  (Deep Xi) \cite{NICOLSON201944} directly from the noisy speech spectral magnitude. The estimated a \textit{priori} SNR can be flexibly employed in statistical model-based methods, e.g., MMSE short-time spectral amplitude (MMSE-STSA) estimator \cite{Y.Ephraim1984}, Log-spectral amplitude MMSE (MMSE-LSA) estimator\cite{Y.Ephraim1985}, and the square-root Wiener filter (SRWF) \cite{1979enhancement}. In this study, the performance evaluation is conducted within the Deep Xi speech enhancement framework.

For speech enhancement task, MLPs-based methods capture temporal information utilizing a context window. However, MLPs are not able to learn the long-term dependencies inherent in noisy speech. In order to capture the long-range dependencies of noisy speech, Chen \textit{et al.} \cite{chen2017long} proposed to use a recurrent neural network (RNN) with four hidden long short-term memory (LSTM) \cite{hochreiter1997long} layers to estimate the IRM. The LSTM model has been shown to generalize to unseen speakers well and significantly outperforms MLP-based models. While able to model the long-range dependencies of speech signal, LSTM-based models exhibit several drawbacks. The high latency and high training complexity of the LSTM model significantly limits its applicability. For RNNs, the other major issue is the vanishing gradients problem, which causes RNNs to be difficult to train.


Over the last decade, convolutional neural networks (CNNs) have gained a considerable amount of success on computer vision and image classification tasks. Recently, CNNs have made great progress in sequence modeling tasks, e.g., speech recognition \cite{CNNASR2014convolutional}. The introduction of causal dilated convolutional units \cite{Dilated2015multi} enables a CNN to obtain an exponentially large receptive field. The dense (DenseNet) \cite{Dense2017} and residual connections (ResNet) \cite{ResNet2016deep} of CNNs allow for both very deep networks and a very long effective history. 
The temporal convolutional network (TCN) exploits ResNet incorporating dilated causal convolutional units, which has allowed CNNs to outperform LSTM models across a diverse range of sequence modeling tasks \cite{TCN2018empirical}. The TCN is able to process the consecutive frames in parallel to significantly speed up the evaluation process, while avoiding the complexity of training a recurrent architecture.


The Inception models \cite{Inception2015going, Inception2016rethinking, Inception2017inception} are successful multi-branch CNN architectures, and an important common property is the split-transform-aggregate design, which enable strong representation power at low computational complexity.
Inspired by the success of Inception models, we investigated the split-transform-aggregate technique for speech enhancement in the proposed mutli-branch temporal convolutional network (MB-TCN). Motivated by the TCN, the MB-TCN incorporates 1-D causal dilated CNN and residual learning for capturing long-term temporal contextual information. 
Experimental results show that the proposed MB-TCN is able to provide more excellent performance and higher parameter efficiency than several advanced networks, i.e., ResLSTM, TCN, and DenseNet. 
Moreover, we also find that our proposed model substantially outperforms several state-of-the-art deep learning based speech enhancement methods.


The remainder of this paper will be structured as follows. In Section II, we give a description of the Deep Xi speech enhancement framework. In Section III, the proposed model is described in detail. To demonstrate the superiority of our model, we conduct the experiments on two datasets (one is ours and one is publicly available dataset used in many previous works). The experimental setup and results on these two datasets are given in Section IV and Section V, respectively. In Section VI, we provide the conclusions and discussions. 

\section{Deep Xi speech enhancement framework}

\subsection{Problem Formulation}
In the time-domain, the noisy-speech signal, $x[n]$, is given by
\begin{equation} \label{equ:b}
x[n] = s[n] + d[n],
\end{equation}
where $s[n]$ and $d[n]$ denote the clean-speech and uncorrelated additive noise, respectively, and $n$ denotes the discrete-time index. The noisy speech, $x[n]$, is then analysed frame-wise using the short-time Fourier transform (STFT):
\begin{equation}
X[l,k] = S[l,k] + D[l,k],
\end{equation}
where $X[l,k]$, $S[l,k]$, and $D[l,k]$ denote the complex-valued STFT coefficients of the noisy speech, the clean speech, and the noise, respectively, for time-frame index $l$ and discrete-frequency index $k$. In polar form, the noisy-speech STFT coefficient is expressed as $X[l,k] = |X[l,k]|e^{j\phi[l,k]}$, where $|X[l,k]|$ and $\phi[l,k]$ denote the noisy-speech short-time spectral magnitude and phase components, respectively. The \textit{a priori} SNR, $\xi[l,k]$, and the \textit{a posteriori} SNR, $\gamma[l,k]$, are defined as follows:
\begin{equation} \label{equ:c}
\xi[l,k] = \frac{\lambda_{s}[l,k]}{\lambda_{d}[l,k]}, \quad  \quad \gamma[l,k] = \frac{|X[l,k]|^2}{\lambda_{d}[l,k]},
\end{equation}
where $\lambda_{s}[l,k]=E\{|S[l,k]|^{2}\}$ and $\lambda_{d}[l,k] = E\{|D[l,k]|^{2}\}$ denote the spectral variances of speech and noise, respectively ($E\{\cdot\}$, is the expectation operator). 

For statistical model-based speech enhancement methods, the estimate of the clean-speech magnitude, $|\widehat{S}[l,k]|$, is obtained using a gain function:
\begin{equation}
    |\widehat{S}[l,k]| = G\left(\xi[l,k], \gamma[l,k]\right)|X[l,k]|.    
\end{equation}
The enhanced speech is then synthesized from the enhanced-speech magnitude spectrum and the noisy-speech phase spectrum using the inverse STFT followed by the overlap-add method \cite{1163353}. The gain function depends on the assumed statistical models of the clean speech and noise and on the criterion that is optimized for. Based on the minimum mean-square error (MMSE) criterion and Gaussian distributions for the clean-speech and noise components, the gain functions of three widely used speech estimators, the SRWF \cite{1979enhancement}, the MMSE-STSA estimator \cite{Y.Ephraim1984}, and the MMSE-LSA estimator \cite{Y.Ephraim1985} are given as 

\begin{equation}
G_{\mathrm{SRWF}}[l,k]=\sqrt{\frac{\xi[l, k]}{\xi[l, k]+1}},
\end{equation}
\begin{equation}
\begin{aligned}  & G_{\mathrm{MMSE}-\mathrm{STSA}}[l,k] =\frac{\sqrt{\pi}}{2} \frac{\sqrt{v[l, k]}}{\gamma[l, k]} \exp \left(\frac{-v[l, k]}{2}\right) \\ & \cdot\left((1+v[l, k]) I_{0}\left(\frac{v[l, k]}{2}\right)+v[l, k] I_{1}\left(\frac{v[l, k]}{2}\right)\right) \end{aligned},
\end{equation}
\begin{equation}
G_{\mathrm{MMSE}-\mathrm{LSA}}[l,k]=\frac{\xi[l, k]}{\xi[l, k]+1} \exp \left\{\frac{1}{2} \int_{v[l, k]}^{\infty} \frac{e^{-t}}{t} d t\right\},
\end{equation}
where $I_{0}(\cdot)$ and $I_{1}(\cdot)$ are the modified Bessel function of zero and first order, respectively, and $v[l,k] = \xi[l, k] \cdot \gamma[l, k] / (\xi[l, k]+1)$. It can be seen that the gain functions depend on two parameters, the a \textit{priori} SNR and the \textit{potseriori} SNR. 

\subsection{Mapped \textit{a priori} SNR training target}

As the enhancement performance of the aforementioned statistical models are predominantly affected by the accuracy of the a \textit{priori} SNR estimate, the training target in Deep Xi framework\footnote{Deep Xi is available at: \url{https://github.com/anicolson/DeepXi}.} is the mapped \textit{a priori} SNR, as described in \cite{NICOLSON201944}. The mapped \textit{a priori} SNR is a mapped version of the instantaneous \textit{a priori} SNR. For the instantaneous case, the clean speech and noise of the noisy speech in (1) are known completely. This means that $\lambda_s[l,k]$ and $\lambda_d[l,k]$ in (3) can be replaced with the squared magnitude of the clean speech and noise spectral components, respectively.

In \cite{NICOLSON201944}, the instantaneous \textit{a priori} SNR (in dB), $\xi_{\rm dB}[l,k]=10\log_{10}(\xi[l,k])$, was mapped to the interval $[0, 1]$ in order to improve the rate of convergence of the used stochastic gradient descent algorithm. The cumulative distribution function (CDF) of $\xi_{\rm dB}[l,k]$ was used as the map. It can be seen in  \cite[Fig. 2]{NICOLSON201944} that the distribution of $\xi_{\rm dB}$ for a given frequency component follows a normal distribution. It was thus assumed that $\xi_{\rm dB}[l,k]$ is distributed normally with mean $\mu_k$ and variance $\sigma^2_k$: $\xi_{\rm dB}[l,k]\sim\mathcal{N}(\mu_k,\sigma^2_k)$. The map is given by
\begin{equation} \label{equa}
\bar{\xi}[l,k] = \frac{1}{2}\Bigg[1 + \textrm{erf}\Bigg( \frac{\xi_{\rm dB}[l,k] - \mu_k}{\sigma_k\sqrt{2}} \Bigg) \Bigg],
\end{equation}
where $\bar{\xi}[l,k]$ is the mapped \textit{a priori} SNR. Following \cite{NICOLSON201944}, the statistics of $\xi_{\rm dB}[l,k]$ for each noisy speech spectral component are found over a sample of the training set.\footnote{The sample mean and variance of $\xi_{\rm dB}[l,k]$ for each noisy speech spectral component were found over $1\,250$ noisy speech signals created from the clean speech and noise training sets (Section \ref{sec:c}). 250 randomly selected (without replacement) clean speech recordings were mixed with random sections of randomly selected (without replacement) noise recordings. Each of these were mixed at five different SNR levels: -5 to 15 dB, in 5 dB increments.} During inference, $\hat{\xi}[l,k]$ is found from $\hat{\xi}_{\text{dB}}[l,k]$ as follows:
\begin{equation}
\hat{\xi}[l,k] = 10^{(\hat{\xi}_{\text{dB}}[l,k]/10)},
\end{equation}
where the \textit{a priori} SNR estimate in dB is computed from the mapped \textit{a priori} SNR estimate as follows:
\begin{equation}
\hat{\xi}_{\rm dB}[l,k] = \sigma_k\sqrt{2}\textrm{erf}^{-1}\big(2\hat{\bar{\xi}}[l,k] - 1\big) + \mu_k.
\end{equation}

\section{Proposed Model}

\subsection{1-D causal dilated convolutional units}

For the proposed network architecture, causal dilated convolutional units are employed for speech enhancement. A causal system has no leakage of information from future time-steps when inferring from the current time-step. For conventional convolutional units (a dilation rate of 1), an extremely deep network or large size kernels are required to build a large temporal receptive field. However, this introduces two typical issues: the vanishing gradient problem and an increase in computational complexity. In \cite{Dilated2015multi}, dilated convolutional units were firstly proposed to achieve a large receptive field for a semantic segmentation task. Its high performance was due to its ability to form a large receptive field, while consuming significantly fewer parameters. 

\begin{figure}[htbp]
\centering
\includegraphics[width=0.48\textwidth]{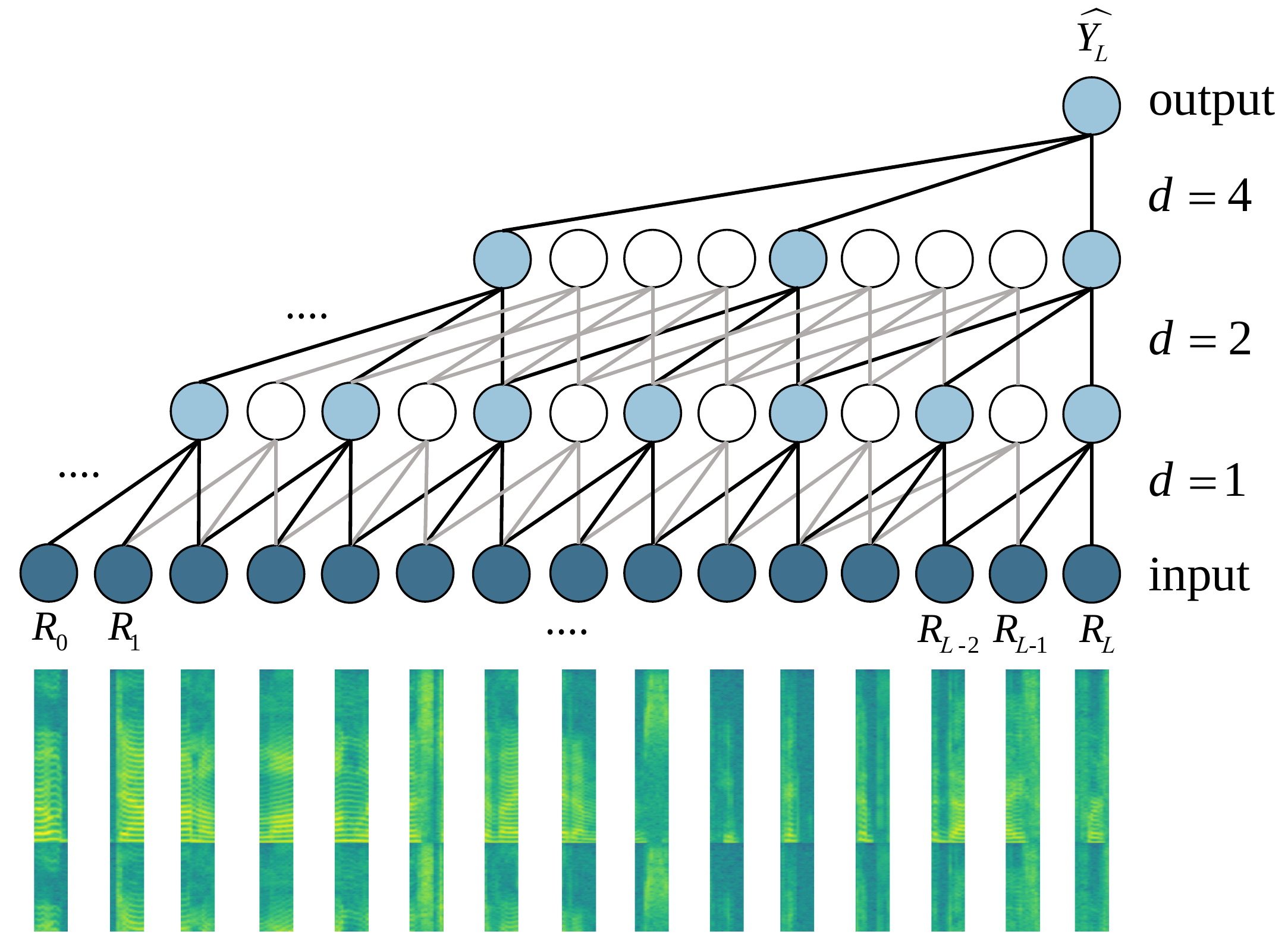}
\caption{An example of 1-D dilated causal convolution with the kernel size $k =3$ and the dilation rate $d = 1,2,4$. This forms a large receptive field over the input.}
\label{fig:c}
\end{figure}

Formally, a 1-D discrete dilated convolution operator $*_{d}$, which convolves a sequence input $\mathbf{x} \in \mathbb{R}^{L}$ with a kernel $f \in \mathbb{R}^{W}$, is represented as
\begin{equation}
    (\mathbf{x}*_{d}f)(l) = \sum_{w=0}^{W-1} f(w)\mathbf{x}(l-d\cdot w)
\end{equation}
where $d$ is the dilation rate and $W$ denotes the kernel size. As a special case, dilated convolution with dilation rate $d=1$ is equivalent to regular convolution. Fig. 1 illustrates an example of 1-D dilated causal convolution with kernel size $k=3$ and dilation factor $d = 1, 2, 4$. As shown in Fig. 1, the receptive field of the network grows exponentially when the dilation factor $d$ increases exponentially with the depth of the network. It enables dilated causal convolutions to capture extremely long-term temporal contextual information using deep networks.

\begin{figure}[htbp]
\centering
\begin{subfigure}[b]{0.24\textwidth}
\centering
\includegraphics[scale=0.85]{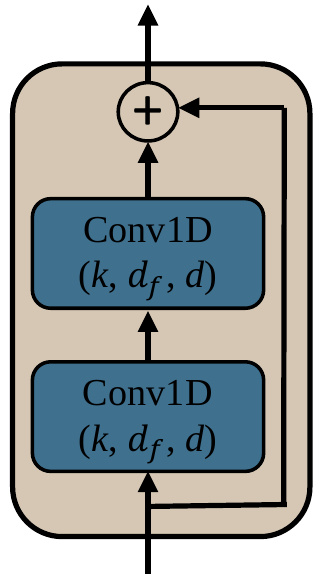}
\caption{}
\end{subfigure}
\begin{subfigure}[b]{0.24\textwidth}
\centering
\includegraphics[scale=0.85]{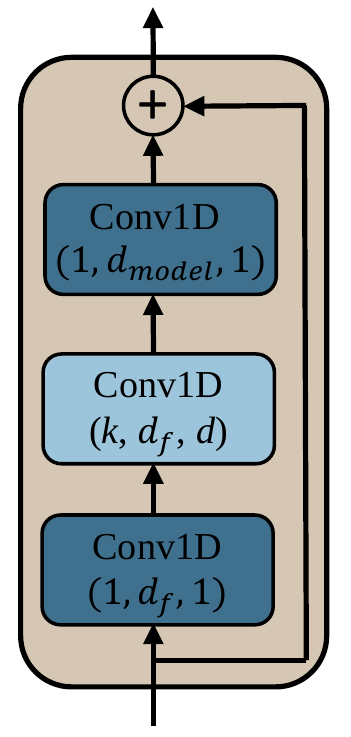}
\caption{}
\end{subfigure}
\caption{The illustration of two commonly used residual blocks of (a) basic structure and (b) bottleneck structure, where $\bigoplus$ represents the element-wise summation operation. The kernel size, output size, and dilation rate for each convolutional unit is denoted as $(\textbf{kernel}~ \textbf{size},~\textbf{output}~\textbf{size},~\textbf{dilation}~\textbf{rate})$.}
\label{fig:c}
\end{figure}

\subsection{Residual Connections}

In addition to the dilation rate and the kernel size, the depth of the model also significantly impacts the temporal receptive field size. While increasing the depth of the model will increase the temporal receptive field size, it also increases the vanishing gradient problem. In \cite{ResNet2016deep}, He \textit{et. al} introduce identity shortcut connections to design a deep residual learning framework to address the vanishing gradient problem. Residual learning has been demonstrated to be an effective way to train very deep networks. In the proposed model, we therefore employ residual blocks in order to facilitate the training of a deep network. Fig. 2(a) and (b) depict basic and bottleneck 1-D residual blocks, respectively. The identity shortcut connections are realised by adding the input of the block to the output of the last convolutional unit.


\subsection{Multi-Branch TCN}

Deep-Xi-MB-TCN is shown in Fig. 3, and is described from input to output as follows. The input to the MB-TCN is the noisy speech magnitude spectrum for the $l^{th}$ frame, $\textbf{R}_l$. The input is first transformed by $\textbf{FC}$, a fully-connected layer of size $d_{model}$ that includes layer normalisation followed by the rectified linear unit (ReLU) activation function. The $\textbf{FC}$ layer is followed by $N$ multi-branch TCN blocks, where $n=1,2,\ldots,N$ is the block index.


\begin{figure}[htbp]
\centering
\includegraphics[width = 0.49\textwidth]{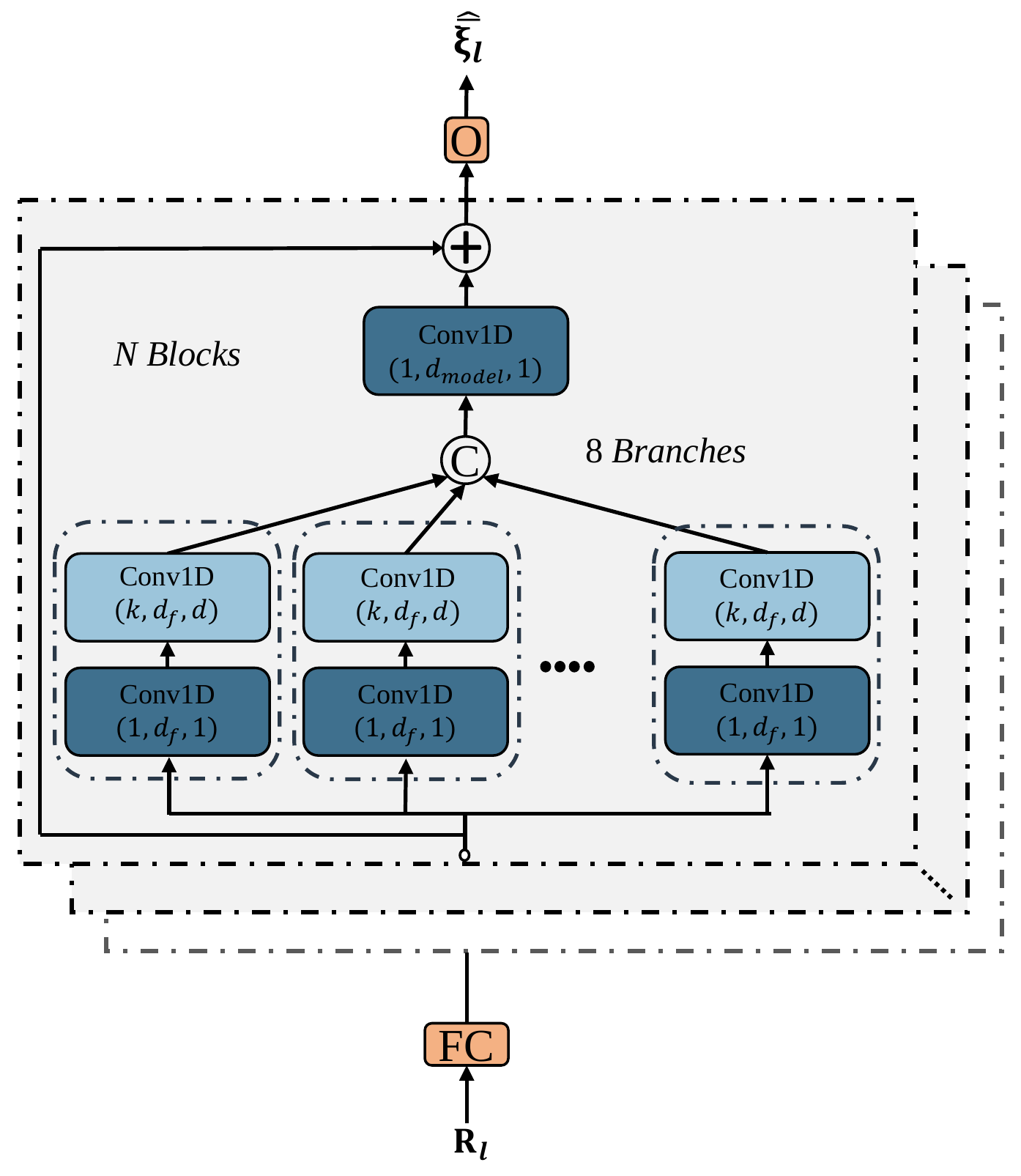}
\caption{Proposed multi-branch TCN (MB-TCN) architecture. It is composed of a fully-connected first layer, \textbf{FC}, followed by \textit{N} multi-branch TCN blocks, and then a fully-connected output layer, \textbf{O} that employs sigmoid units. $\copyright$ represent the concatenation operation. The kernel size, output size, and dilation rate for each convolutional unit is denoted as $(\textbf{kernel}~ \textbf{size},~\textbf{output}~\textbf{size},~\textbf{dilation}~\textbf{rate})$.}
\label{fig:c}
\end{figure}

\subsubsection{Split-transform-aggregate strategy} 
Inception models \cite{Inception2015going, Inception2016rethinking, Inception2017inception},  are successful multi-branch architectures, and their effectiveness has been proven by various computer vision tasks. The most important design of the Inception models is the split-transform-aggregate strategy: the input is split into several low-dimensional representations, transformed by each customized branch network, and aggregated by
concatenation. The split-transform-aggregate design enables the Inception models to demonstrate strong representation power at a low computational complexity. Inspired by the success of Inception models, the design of MB-TCN exploits split-transform-aggregate network architecture that incorporates 1-D dilated convolutional units for speech enhancement task. Each multi-branch residual block includes branching (8 branches), concatenating operations, and residual connection. 

\subsubsection{Topology of each branch}
The architecture for each branch network in Inception models is carefully designed to obtain good performance for a specific task. However, it is difficult to adapt the Inception models to new tasks as many hyper-parameters need to be designed. To improve the extendibility, Xie \textit{et al,} \cite{ResNext2017} proposed to utilize the same topology for all paths to simply the realization. Motivated by this, the residual block shares the same topology for all branches, including two 1-D causal dilated convolutional units (with same size and dilation rate), where each convolutional unit is pre-activated by layer normalisation \cite{LN2016layer} followed by the ReLU activation function. The kernel size, output size, and dilation rate for each convolutional unit is denoted in Fig. 3 as $(\textbf{kernel}~\textbf{size},~\textbf{output}~\textbf{size},~\textbf{dilation}~\textbf{rate})$. The first convolutional unit on each path has a kernel size of 1, whilst the second convolutional unit has a kernel size of $k$. The first and second convolutional units have an output size of $d_f$. The first convolution unit compress the input to a low-dimensional embedding. The second convolutional unit has a dilation rate of $d$, providing the capability of capturing the long-term contextual information. As in \cite{TaeNet2018tasnet}, the dilation rate $d$ is cycled as the block index $n$ increases: $d = 2^{((n - 1) \; \textrm{mod} \; (\log_{2}(D) + 1))}$, where $\textrm{mod}$ is the modulo operation, and $D$ is the maximum dilation rate. Such stacked residual blocks support the exponential expansion of the receptive field without loss of input resolution, which allows for capturing long-term effective history. 

The feature-maps produced by different branch networks are aggregated by concatenation. Then the aggregate feature-maps is processed with a 1-D convolutional unit and the output of a multi-branch residual network is produced by using a identity residual connection. This convolutional unit has an output size of $d_{model}$ and is also pre-activated by layer normalisation followed by the ReLU activation function \cite{Relu2010}.

The last block is followed by the output layer, $\textbf{O}$, which is a fully-connected layer with sigmoidal units. The $\textbf{O}$ layer estimates the mapped \textit{a priori} SNR for each spectral component of the $l^{th}$ time-frame, $\hat{\bar{\pmb{\xi}}}_l$.
The following hyperparameters were chosen for the network, as a compromise between training time and performance: $d_{model}=256$ and  $d_f=64$. As in \cite{DBLP:journals/corr/KalchbrennerESO16}, $k$ is set to $3$, and $D$ is set to $16$. MB-TCN with size of 1.05, 1.43, and 1.66 million parameters are formed by cascading the 12, 17, and 20 multi-branch residual blocks. The MB-TCNs with 12, 17, and 20 multi-branch TCN blocks accumulate a receptive field of 2.1, 3.1, and 4 seconds, respectively, over past and present time-frames



\begin{figure*}[!bp]
\centering
\includegraphics[scale=0.75]{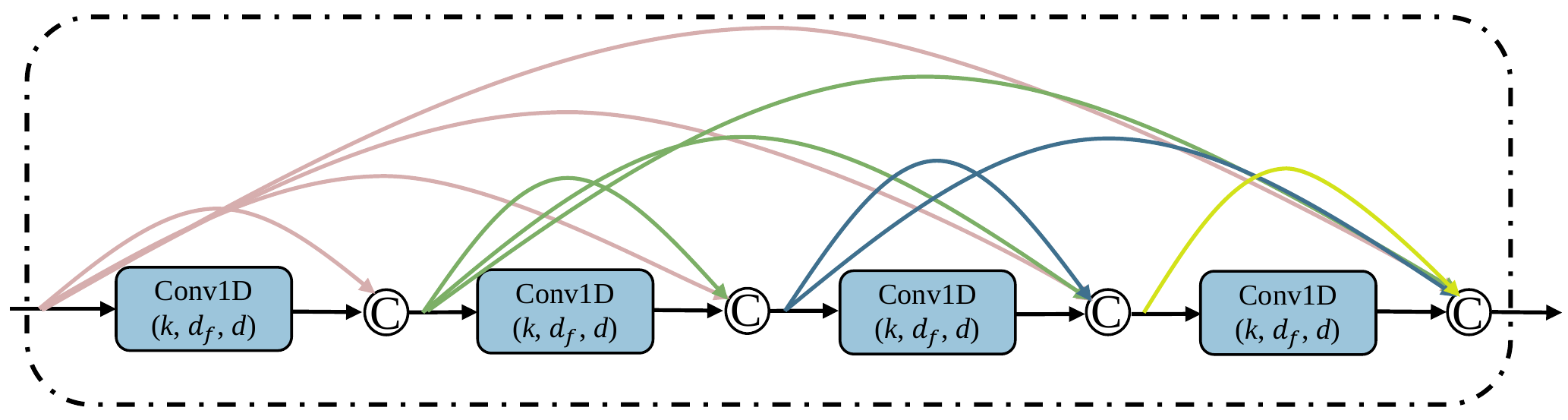}
\caption{The illustration of DenseNet block. Each dense block includes four 1-D causal dilation convolution units. The kernel size, output size, and dilation rate for each convolutional unit is denoted as $(\textbf{kernel}~ \textbf{size},~\textbf{output}~\textbf{size},~\textbf{dilation}~\textbf{rate})$.}
\label{fig:c}
\end{figure*}

\section{Experimental Setup} \label{sec: exp}

\subsection{Signal Processing} \label{secb_short}

The Hamming window function is used for spectral analysis and synthesis \cite{237532, Huang:2001:SLP:560905, 4682553}, with a frame-length of 32 ms ($512$ time-domain samples) and a frame-shift of 16 ms ($256$ time-domain samples). The a \textit{priori} SNR was estimated from the 257-point single-sided noisy speech magnitude, which included both the DC frequency component and the Nyquist frequency component. The \textit{a posteriori} SNR for MMSE-based speech estimators is estimated from the \textit{a priori} SNR estimate \cite{NICOLSON201944}: $\hat{\gamma}[l,k]=\hat{\xi}[l,k]+1$.

\subsection{Baseline Models}

In our experiments, we compare the proposed MB-TCN with the following network architectures (baselines) tasked with estimating the a \textit{priori} SNR for MMSE-based methods to speech enhancement (Deep Xi framework):

\textbf{ResLSTM}: As baselines, we use three residual LSTMs (ResLSTMs) composed of 4, 5, and 6 residual blocks, and the memory cell sizes for each ResLSTM are 170, 188, and 200, respectively. The numbers of parameters for the three ResLSTMs are 1.02, 1.51, and 2.03 million, respectively.

\textbf{TCN-BC}: The TCN-BC models \cite{TCN2018empirical} are formed with basic residual blocks that incorporates 1-D dilated convolutional units. As shown in Fig. 2(a), each residual block contains two 1-D causal dilated convolution units with an output size of $d_{f} = 64$, where each convolution unit is pre-activated by a layer normalisation followed by the ReLU activation function. The kernel size of each convolution units is $k = 3$. The dilation rate $d$  in each block is cycled from 1 to 16 (increasing by power of 2). By cascading 40, 60, and 80 basic residual blocks, we build three TCN-BC models with 1.03, 1.53, and 2.03 million parameters, respectively, as baselines. 


\textbf{TCN-BK}: The TCN-BK models are built with bottleneck residual blocks that incorporates 1-D dilated convolutional units. As shown in Fig. 2(b), each residual block contains three 1-D causal dilated convolution units, where each convolution unit is pre-activated by a layer normalisation followed by the ReLU activation function. The first and third convolution units in a bottleneck residual block have a kernel size of 1, whilst the second unit has a kernel size of $k=3$. The dilation rate $d$ is cycled from 1 to 16 (increasing by power of 2). By cascading 20, 30, and 40 bottleneck residual blocks, we build three TCN-BK models size of 1.05, 1.51, and 1.98 million parameters, respectively.


\textbf{DenseNet}: Fig. 4 shows that each dense block consist of four dilated causal convolution units with a kernel size of $k =3$, where each convolution unit is pre-activated by a layer normalisation followed by the ReLU activation function. The output size of each convolution unit is $d_{f} = 24$. The dilation rate $d$ is cycled from 1 to 16 (increasing by power of 2). As baselines, DenseNets of sizes 0.97, 1.48, and 2.10 million parameters are formed by cascading 7, 9, and 11 dense blocks, respectively.

In addition to the aforementioned models which are evaluated in Deep Xi framework, the proposed enhancement system is also compared with two widely known deep learning speech enhancement methods, LSTM-IRM proposed in \cite{chen2017long} and bidirectional LSTM (BLSTM)-IRM. For these two methods, ideal ratio mask (IRM) is used as training target and 23 consecutive input frames (11 past frames, 1 current frames, and 11 future frames) are concatenated into a feature vector that used as the input of the network to estimate current frame of the mask. The network used in LSTM-IRM is composed of four LSTM layers, each with 1024 units, and a fully connected layers with 257 sigmoid units. The network used in BLSTM-IRM consists of four bidirectional LSTM layers, each with 512 units, and a fully connected layers with 257 sigmoid units. 

\subsection{Database} \label{sec:c}

\textbf{Training Set}: Here, we give a detailed description of the clean speech and noise recordings used for training models in this experiment. For the clean speech recordings, we use the \textit{train-clean-100} set from the Librispeech corpus as training set \cite{7178964}, which includes $28\,539$ utterances spoken by 251 speakers. To conduct cross-validation experiments, 1000 clean speech recordings are randomly selected from the training set to construct the validation set. The employed noise recordings in the training set are taken from the following datasets: the QUT-NOISE dataset \cite{Dean2010}, the Nonspeech dataset \cite{Nonspeech}, the Environmental Background Noise dataset \cite{7472068, 7590807}, and the noise set from the MUSAN corpus \cite{DBLP:journals/corr/SnyderCP15}. This gives a total of $1\,295$ noise recordings. All clean speech and noise recordings are single-channel, with a sampling frequency of 16 kHz (recordings with a sampling frequency higher than 16 kHz are downsampled to 16 kHz). A description of how the noisy speech signals are generated from the clean speech and noise recordings is given in the next subsection.


\textbf{Test Set}: 
For testing, recordings of four different noise sources are employed to form the test set. Two of the four noise recordings are of the real-world non-stationary noise sources, which includes \textit{street music} noise (recording no. $26\,270$) from the Urban Sound dataset \cite{Salamon:UrbanSound:ACMMM:14}, and \textit{voice babble} noise from the RSG-10 noise dataset \cite{steeneken1988description}. Another two of the four noise recordings of real-world coloured noise sources, including \textit{factory} and \textit{F16} noises from the RSG-10 noise dataset \cite{steeneken1988description}. We randomly choose 10 clean speech recordings (without replacement) from the TSP speech corpus \cite{kabal2002tsp} for each of the four noise recordings. To construct the noisy speech signals, we mix the clean speech signals with a random section of the noise recording at five different SNR levels: ranging from -5 to 15 dB with a step of 5 dB. This constructed a test set of 200 noisy speech signals. All the noisy speech signals were single channel with a sampling frequency of 16 kHz.

\input{pesq_stoi_dataset_1.tex}

\subsection{Training Details} \label{secd}
The training details for all the models (MB-TCN and baseline models) are described as follows:  
\begin{itemize}
  \setlength\itemsep{0.5pt}
  \item Cross-entropy as the loss function.
  \item We employ the \textit{Adam} optimizer \cite{DBLP:journals/corr/KingmaB14} with $\beta_{1} = 0.9$, $\beta_{2} = 0.999$, and a learning rate of 0.001 for gradient descent optimisation.
  \item Gradients are clipped between $[-1,1]$.
  \item The selection order for the clean speech recordings is randomised for each epoch.
  \item The number of training examples in an epoch is equal to the number of clean speech files in the training set ($27\,539$). A total of 105 epochs is used to train all the models except the ResLSTM models, and the two IRM estimators (LSTM-IRM and BLSTM-IRM) replicated from \cite{chen2017long}, where 10 epochs were used. The number of training examples in an epoch is equal to the number of clean speech files in the training set ($27\,539$).
  \item A mini-batch size of $10$ noisy speech signals. For each mini-batch, all samples are padded with zero to keep the same time steps as the longest sample.
  \item The noisy speech signals are created as follows: each clean speech recording selected from the mini-batch is mixed with a random section of a randomly selected noise recording at a randomly selected SNR level (-20 to 30 dB with a step of 1 dB).

\end{itemize}

\subsection{Evaluation Metrics}

In our experiments, two objective metrics are used to evaluate the objective quality and intelligibility of the enhanced speech. The wideband perceptual evaluation of speech quality (Wideband PESQ) metric \cite{WidePESQ} is used to obtain the mean opinion score of the objective speech listening quality. The PESQ score is typically between 1 and 4.5, with a higher score implying better speech quality. The short-time objective intelligibility (STOI) \cite{5495701} is used to evaluate the objective speech intelligibility. The STOI score is typically between 0 and 1, with a higher score indicating better speech intelligibility. In addition, we also show the segmental SNR improvements (SSNR+) over all of the tested conditions.



\section{Results}

The wideband PESQ scores of the enhanced speech signals produced by each of the networks (ResLSTM, DenseNet, TCN-BC, TCN-BK, and MB-TCN) within Deep Xi framework are listed in Table \ref{taba}. The a \textit{priori} SNR is estimated by each network and then employed in the square-root wiener filtering (SRWF) estimator to obtain the enhanced speech. The maximum PESQ score for each condition and each parameter size is highlighted in boldface. From Table \ref{taba} it can be observed that MB-TCN performs best in terms of objective speech quality scores for most of the tested conditions. The performance superiority of MB-TCN is demonstrated as MB-TCN with a parameter size of 1.05 million provides 0.21 PESQ improvement over TCN-BK with same parameter size for \textit{F16} at 15 dB. The MB-TCN also shows a superiority in terms of model size as the MB-TCN with a parameter size of 1.43 million attains 0.11 PESQ improvement over TCN-BK with a parameter size of 1.51 million for \textit{Factory} at 15 dB.

In Table \ref{tabb} we present the STOI scores of the enhanced speech signals produced by each of the models. From Table \ref{tabb} it is seen, similarly to Table \ref{taba}, that MB-TCN is able to attain the highest STOI score under most of the tested conditions. MB-TCN demonstrates its superiority by the evaluation results: for \textit{voice babble} at -5, 0, and 5dB, MB-TCN with a parameter size of 1.66 million obtains 5\%, 4.9\%, and 2\% STOI improvements over TCN-BK with 1.98 million parameters, respectively. Compared to LSTM-IRM and BLSTM-IRM, the enhanced speech signals produced by MB-TCN in Deep Xi framework have higher speech quality and intelligibility scores under almost all noisy conditions except for \textit{ Voice Babble} at -5 dB where BLSTM-IRM achieves a higher STOI score. The average segmental SNR improvement (SSNR+) over all of the tested conditions is shown in Fig. 5. It is observed that our proposed MB-TCNs produces higher SSNR+ than the baselines at multiple parameter sizes.

\begin{figure}[htbp]
\centering
\includegraphics[width = 0.48\textwidth]{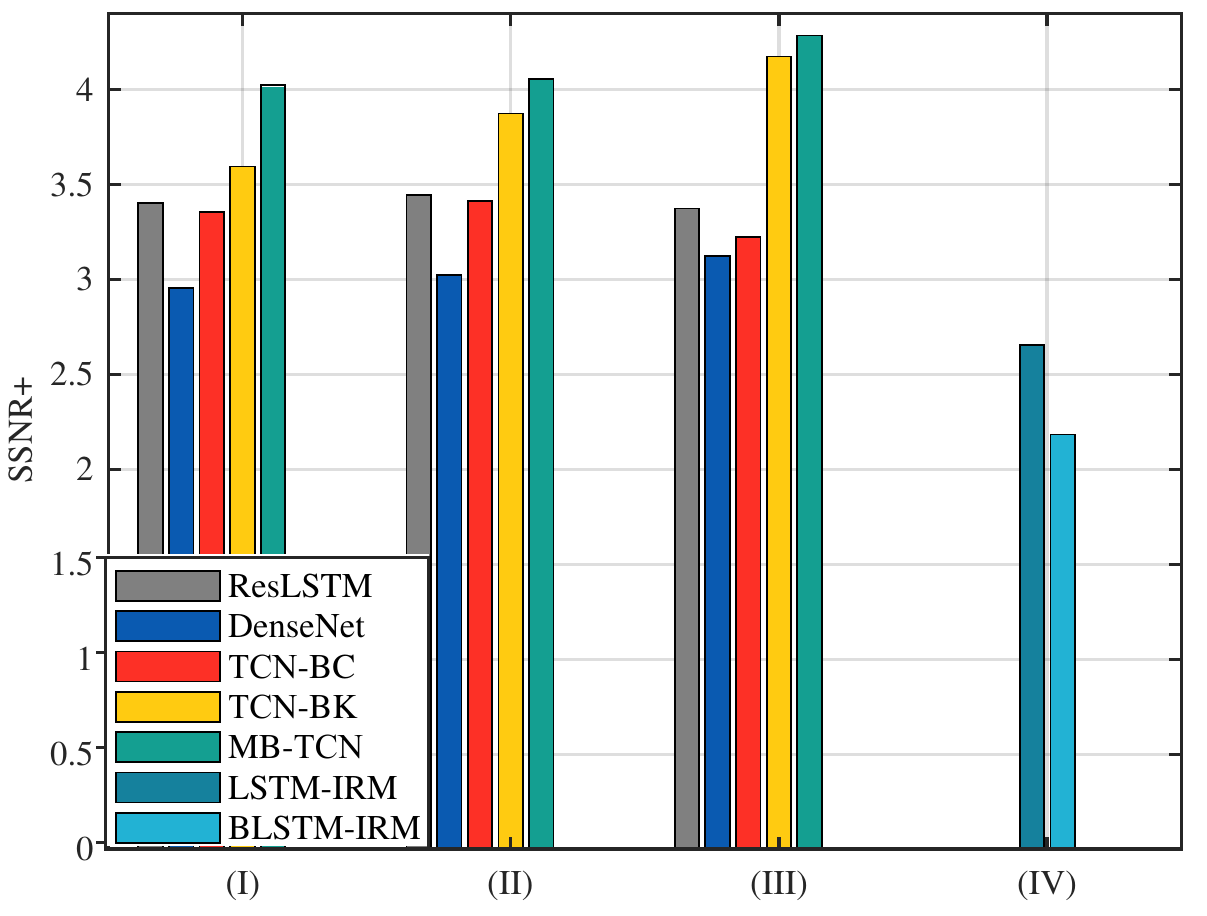}
\caption{The average SSNR+ over all of the tested conditions. The models in (I)-(III) are ResLSTM (1.02, 1.51, and 2.03 M), DenseNet (0.97, 1.48, and 1.94 M), TCN-BC (1.03, 1.53, and 2.03 M), TCN-BK (1.05, 1.51, and 1.98 M), and MB-TCN (1.05, 1.43, and 1.66 M). The models in (IV) are LSTM-IRM (53.9 M) and BLSTM-IRM (39.0 M).} 
\label{fig:d}
\end{figure}

Fig. 6 illustrates the magnitude (log scale) spectrograms of an enhanced speech utterance produced by different models. The enhanced spectrograms shown in Fig. 6(c)-(f) and Fig.6 (i) are produced in Deep Xi-SRWF framework. It can be seen that our proposed MB-TCN achieve the best noise suppression, while preserving the formant information of the clean speech. The spectrograms (g)-(h) generated by LSTM-IRM and BLSTM-IRM show much more residual noise components compared to the enhanced spectrograms by other models. As shown in Fig. 6(d) and (f), DenseNet and TCN-BK demonstrate superior noise suppression, but with severe formant information destruction. Both ResLSTM and TCN-BC preserve the formant information, but ResLSTM shows better noise suppression than TCN-BK. The spectrograms exhibit the consistency with the results in Tables \ref{taba}--\ref{tabb} (\textit{voice babble} at -5 dB).

\begin{figure}[htbp]
\centering
\includegraphics[width = 0.5\textwidth]{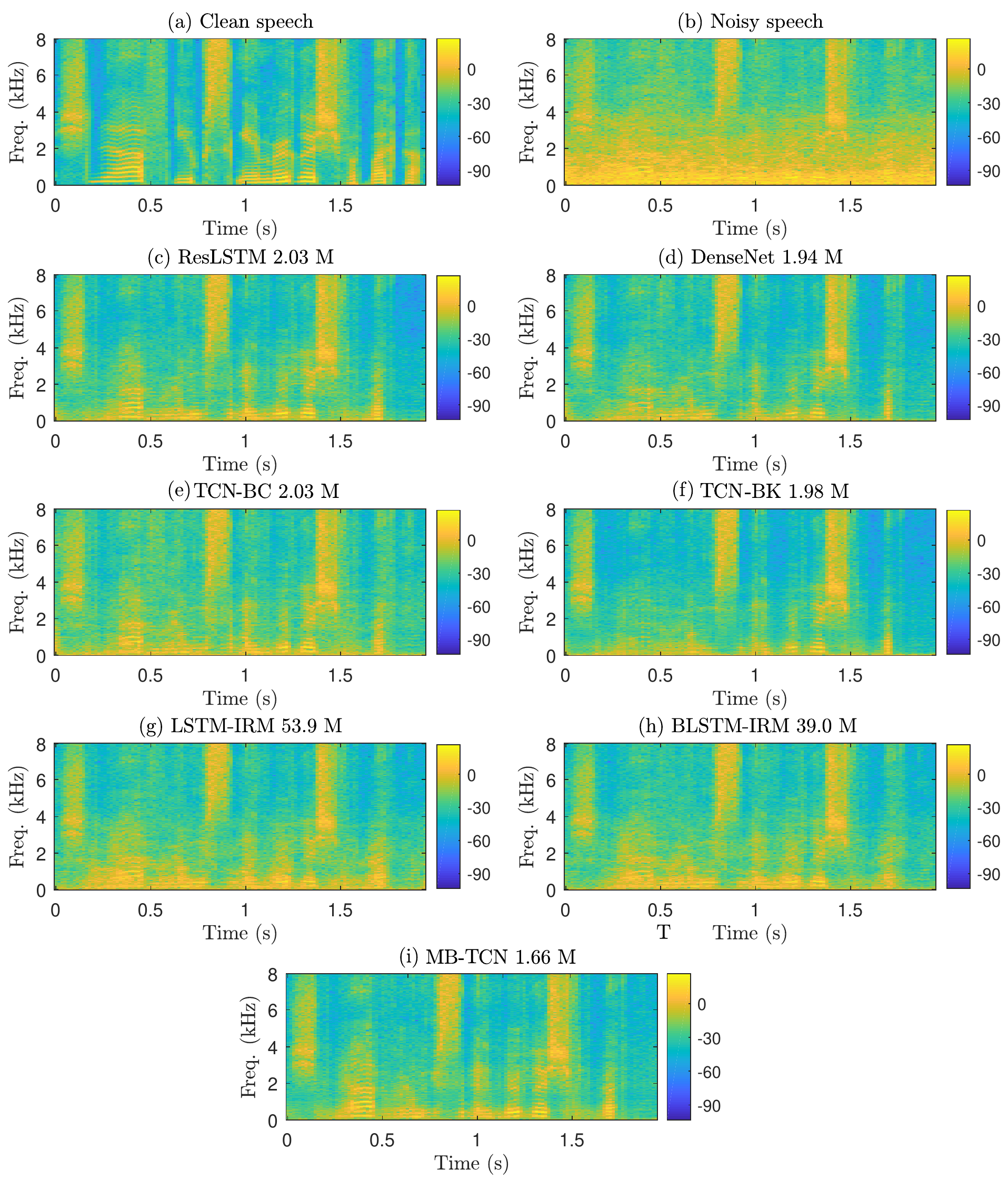}
\caption{Magnitude spectrograms (log scale) of (a) clean speech and (b) noisy speech (clean speech was mixed with \textit{voice babble} at -5 dB). Enhanced speech produced by (c) ResLSTM 2.03 M, (d) DenseNet 1.94 M, (e) TCN-BC 2.03 M, (f) TCN-BK 1.98 M, (g) LSTM-IRM 53.9 M, (h) BLSTM-IRM 39.0 M, and (i) proposed MB-TCN 1.66 M.} 
\label{fig:d}
\end{figure}

\subsection{Parameter Efficiency}

For many real-world speech processing applications, memory and computational resources are considerable constraints. For comparison of parameter efficiency, in Tables \ref{taba} and \ref{tabb} we present the number of trainable parameters in different models. From the numbers in the tables, it can be seen that our proposed MB-TCN exhibits higher parameter efficiency than ResLSTM, DenseNet, TCN-BC, and TCN-BK models. Compared to another two widely known enhancement methods, LSTM-IRM and BLSTM-IRM, the proposed model also demonstrates a significant superiority in terms of parameter efficiency for low-power and low-memory required applications. To be specific, the parameter sizes of LSTM-IRM (53.9 M) and BLSTM-IRM (39.0 M) are around 32.5 times and 23.5 times that of MB-TCN (1.66 M), respectively.

\section{Comparison With Other State-of-the-Art Methods}

In this section, to further demonstrate its superiority, we compare the proposed model with other state-of-the-art methods on a same publicly available dataset. A fair comparison is ensured since all the models are optimized by the authors on the exact same dataset. The brief descriptions for the baseline methods are provided in next subsection.

\input{composite.tex}

\subsection{Baseline Methods} \label{sec:test2_baselin}

In this experiment, we conduct the performance evaluations of the proposed MB-TCN model (in Deep Xi framework) in comparison with the following baseline methods from the literature:
\begin{itemize}
    \item \textbf{Wiener} \cite{wiener1996}, a traditional statistic-based Wiener filtering method that is based on the a  \textit{priori} SNR estimator.
    \item \textbf{SEGAN} \cite{SEGAN2017}, a time-domain speech enhancement model, using generative adversarial networks (GAN) to directly reconstruct the clean waveform from noisy speech waveform. 
    \item \textbf{Wavenet}\cite{wavenet2018}, a non-causal Wavenet-based denosing model, operating on the raw waveform. It employs a regression loss function ($L_{1}$ losses on both the speech waveform and the noise waveform prediction branches).
    \item \textbf{Wave-U-Net} \cite{waveunet2018}, a one-dimensional adaptation of U-Net architecture for time-domain speech enhancement.
    \item \textbf{Deep Feature Loss} \cite{deeploss2018}, also a time-domain denoising model that is trained with a deep feature loss from another acoustic environment classifier network.
    \item \textbf{MMSE-GAN} \cite{MMSE-GAN2018}, a time-frequency (T-F) masking based method, using a modified GAN to predict the clean T-F representation. The objective function includes a GAN objective and an $L_{2}$ loss between the predicted and the clean T-F representation.
    \item \textbf{Metric-GAN} \cite{metricGAN2019}, a T-F masking based method, using a GAN to directly optimize generators based on one or multiple evaluation metric scores to speech enhancement.
    \item \textbf{MDPhD} \cite{MDPhD2018}, a hybrid speech enhancement method of time-domain and time-frequency domain.
\end{itemize}

\subsection{Database}

To make a direct and fair performance comparison, we used the same publicly available dataset \cite{database16} used in several previous works. In this dataset, the clean speech recordings comprise of 30 speakers from the Voice Bank Corpus \cite{VBC2013} -- 28 speakers were chosen for training and the remaining 2 for testing. The noisy speech in training set is synthesized using a mixture of clean speech with 10 types of noise, two of which are artificially generated and 8 real noise recordings are from the Diverse Environments Multi-channel Acoustics Noise Database (DEMAND) \cite{DEMAND013}. With respect to the test set, 20 different noisy conditions are included: 5 distinct types of noise sources from the DEMAND database at one of 4 SNR levels each (2.5, 7.5, 12.5, and 17.5 dB). In total, this produced 842 test samples (approximately 20 different sentences in each condition per test speaker). Both speakers and noise conditions in test set are totally unseen during training process. As in previous methods, the original raw waveforms were downsampled from 48 kHz to 16 kHz for training and testing.


\subsection{Performance Metrics}
In addition to PESQ and STOI metrics, another three composite measures are also exploited to evaluate the enhancement performance of the proposed models and state-of-the-art competitors. The three composite metrics are: 
\begin{itemize}
\item CSIG \cite{2007evaluation}: mean opinion score (MOS) predictor of signal distortion attending only to the speech signal (from 1 to 5).
\item CBAK \cite{2007evaluation}: MOS predictor of the background-noise intrusiveness (from 1 to 5).
\item COVL \cite{2007evaluation}: MOS predictor of the overall speech quality (from 1 to 5).
\end{itemize}
\subsection{Experimental Results}


Table \ref{tab:composite} presents the comparison results of these metrics on the second dataset. For all baseline methods, the best results that have been reported in literature are listed. The missing values in the table are because the results are not reported in the work. Here, the Deep Xi framework using proposed MB-TCN to estimate the a \textit{priori} SNR is integrated into the SRWF, MMSE-STSA, and MMSE-LSA speech estimators. 

\begin{figure}[htbp]
\centering
\includegraphics[width = 0.5\textwidth]{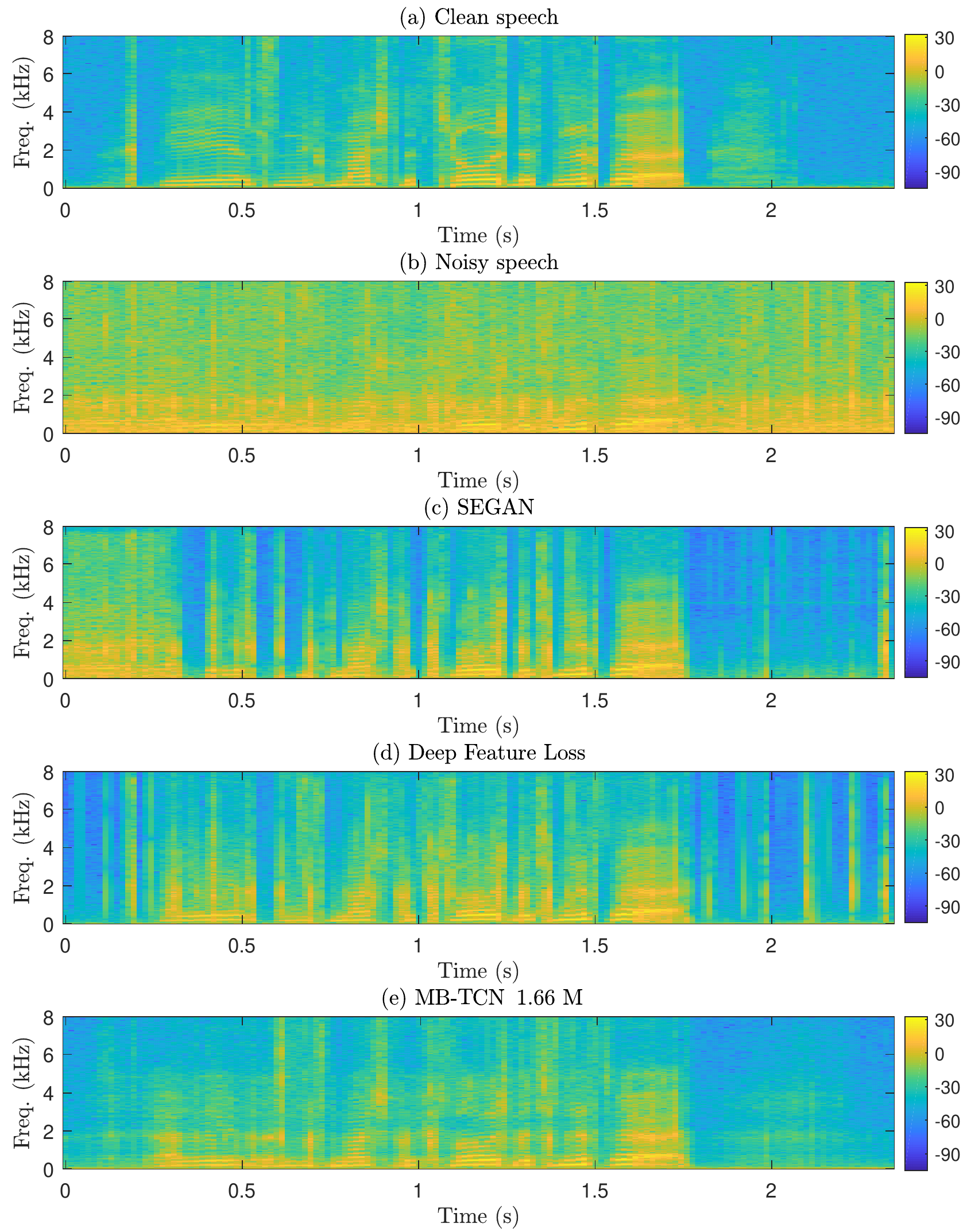}
\caption{Magnitude spectrograms (log scale) of (a) clean speech (female \textit{p}257 uttering sentence 151) and (b) noisy speech (clean speech mixed with \textit{street noise} at 2.5 dB). Enhanced speech produced by (c) SEGAN, (d) Deep Feature Loss, and (e) proposed MB-TCN (Deep Xi-SRWF 1.66 M).} 
\label{fig:d}
\end{figure}

It can be clearly observed in Table \ref{tab:composite} that the MB-TCN outperforms time-domain methods such as SEGAN \cite{SEGAN2017}, Wavenet \cite{wavenet2018}, Wave-U-Net \cite{waveunet2018}, and Deep Feature Loss (DFL) \cite{deeploss2018} in terms of all five measures by a comfortable margin. For example, MB-TCN provides 0.35, 0.08, and 0.37 improvements over DFL for CSIG, CBAK, and COVL, respectively. Our method also shows large performance gain over T-F based methods like MMSE-GAN \cite{MMSE-GAN2018} and MetrciGAN \cite{metricGAN2019}. For example, the MB-TCN provides 0.22, 0.23, 0.17, and 0.08 improvements over MetricGAN for CSIG, CBAK, COVL, and PESQ, respectively. The proposed model also provides 0.41 PESQ improvements and about 1\% STOI improvements over MMSE-GAN. In addition, our model provides great improvement over a hybrid method of time-domain and time-frequency domain, MDPhD \cite{MDPhD2018}. These evaluation results significantly demonstrate the superiority of the proposed model. 

The magnitude (log scale) spectrograms of enhanced speech produced by SEGAN, DFL, and MB-TCN (Deep Xi-SRWF 1.66 M) are shown in Fig. 7(c)-(e), respectively. It can be observed that MB-TCN (e) is able to achieve better trade-off between noise suppression and speech distortion. At the beginning segment of speech (around 0-0.35 s), SEGAN almost exhibits no noise suppression (Fig. 7(c)). In addition, multiple residual noise components can be seen in the spectrograms enhanced by both SEGAN and DFL (Fig. 7(d)).

\subsection{Parameter Efficiency}

As mentioned in Section V-A, the parameter efficiency of models is a considerable constraints for many real-world speech applications.  For this, in Table \ref{tab:composite} we present the the number (in millions) of learnable parameters in different models. In Table \ref{tab:composite}, the numbers of learnable parameters in different models listed are reported value in literature or computed from the code provided by the authors. Additionally, since GAN-based methods need to train both generator and discriminator networks, the listed values represent the number of parameters in generator and discriminator (in bracket), respectively. From the Table \ref{tab:composite} we can find that our proposed model is able to provide higher parameter efficiency than most state-of-the-art (SOTA) methods (SEGAN, Wavenet, Wave-U-Net, Metric-GAN, and MDPhD). Although MMSE-GAN has less parameter than proposed model, the training instabilities of GAN models are still not completely understood. For DFL, an extra feature loss model needs to be pre-trained. One must note that our proposed model has a significant performance improvement accompanied by a modest increasing of number of trainable parameters compared to MMSE-GAN and DFL. In addition, note that we can adjust the parameter efficiency of MB-TCN simply by altering the multi-branch dilated residual blocks. Compared to these SOTA models, the presented model is able to achieve better trade-off between performance and parameter efficiency.

\section{Conclusion and Discussion} \label{sec: cd}

In this study, we have presented an MB-TCN model for monaural speech enhancement. The proposed model utilizes the split-transform-aggregate design, and incorporates the 1-D causal dilated convolutions and identity residual connection. The split-transform-aggregate design demonstrates a strong representation power at low computational complexity. The combination of dilated convolutions and residual learning builds large receptive fields, which enables our proposed model the ability to capture very long effective history information to make a prediction. Specifically, large receptive fields enable model to learn the temporal dynamics of speech very well. The experimental results demonstrate that our proposed model outperforms many other advanced networks, such as ResLSTM, TCN with basic structure (TCN-BC), TCN with bottleneck structure (TCN-BK), and DenseNet. Compared to two widely known deep learning methods, LSTM-IRM and BLSTM-IRM, the MB-TCN in Deep Xi framework shows a significant superiority in term of both performance and parameter efficiency. 

Moreover, the comparison results with many state-of-the-art (SOTA) speech enhancement algorithms also demonstrates that our method is able to provide better enhancement performance in terms of widely used five objective metrics. For low-power and low-memory required applications of deep learning based speech enhancement methods, the number of parameters (parameter efficiency) in models is considerable constraint. It is crucial to achieve an optimal trade-off between parameter efficiency and speech enhancement performance of the model. However, most SOTA baseline models are not able to provide high parameter efficiency. The experimental results demonstrate that our proposed model is able to achieve a better trade-off than many SOTA speech enhancement methods.

\ifCLASSOPTIONcaptionsoff
  \newpage
\fi



\bibliographystyle{IEEEtran}
\bibliography{IEEEabrv,./myreference}
\end{document}

%% file: pesq_stoi_dataset_1.tex
\begin{table*}[ht]
    \centering
    \scriptsize
    \def\arraystretch{1.35}
    \setlength{\tabcolsep}{3.6pt}
    \caption{
    Speech enhancement performance of different networks in terms of wideband PESQ metric. The highest PESQ score obtained at each condition and for each parameter size is highlighted with bold text.
    }
  \begin{tabular}{lc|lllll|lllll|lllll|lllll} 
  \toprule
   \multirow{3}{*}{ \textbf{Network} } & \multirow{3}{*}{\begin{tabular}[c]{@{}l@{}}\textbf{\# params.}\\$\pmb{\times 10^{6}}$ \end{tabular}} & \multicolumn{20}{c}{\textbf{SNR level (dB)}} \\
   
    \cline{3-22} &  & \multicolumn{5}{c|}{\textbf{Voice babble}} & \multicolumn{5}{c|}{\textbf{Street music}} & \multicolumn{5}{c|}{\textbf{F16}} & \multicolumn{5}{c}{\textbf{Factory}} \\
    \cline{3-22}& & {\bf-5}  & {\bf0}   & {\bf5}   & {\bf10}  & {\bf15}        & {\bf-5}  & {\bf0}   & {\bf5}   & {\bf10}  & {\bf15} & {\bf-5}  & {\bf0}   & {\bf5}   & {\bf10}  & {\bf15}  & {\bf-5}  & {\bf0}   & {\bf5}   & {\bf10}  & {\bf15}           \\ 
    \hline
        
Noisy speech & -- & 1.04 & 1.07 & 1.14 & 1.35 & 1.71 & 1.04 & 1.06 & 1.11 & 1.27 & 1.58 & 1.03 & 1.06 & 1.11 & 1.25 & 1.52 & 1.04 & 1.04 & 1.09 & 1.24 & 1.54 \\

\midrule
ResLSTM & 1.02 & 1.08 & 1.19 & 1.43 & 1.91 & 2.44 & 1.08 & 1.18 & 1.37 & 1.70 & 2.14 & 1.12 & 1.27 & 1.54 & 1.87 & 2.28 & 1.06 & 1.22 & 1.49 & 1.87 & 2.32 \\

DenseNet & 0.97 & 1.05 & 1.16 & 1.41 & 1.83 & 2.28 & 1.06 & 1.15 & 1.36 & 1.64 & 2.09 & 1.11 & 1.26 & 1.47 & 1.76 & 2.13 & 1.04 & 1.15 & 1.39 & 1.74 & 2.17\\ 

TCN-BC & 1.03 & 1.07 & 1.18 & 1.42 & 1.87 & 2.34 & 1.09 & 1.20 & 1.43 & 1.75 & 2.21 & 1.13 & 1.31 & 1.57 & 1.89 & 2.29 & 1.07 & 1.23 & 1.50 & 1.89 & 2.35 \\

TCN-BK & 1.05 & 1.08 & 1.23 & 1.53 & 1.92 & 2.37 & 1.10 & 1.24 & 1.49 & 1.80 & 2.24 & 
\textbf{1.16} & 1.36 & 1.60 & 1.88 & 2.25 & 
1.11 & \textbf{1.30} & 1.55 & 1.85 & 2.28\\ 


Prop. MB-TCN & 1.05 & \textbf{1.09} & \textbf{1.25} & \textbf{1.55} & \textbf{2.04} & \textbf{2.57} & \textbf{1.11} & \textbf{1.26} & \textbf{1.52} & \textbf{1.87} & \textbf{2.38} & \textbf{1.16} & \textbf{1.37} & \textbf{1.65} & \textbf{2.03} & \textbf{2.46} & \textbf{1.12} & 1.29 & \textbf{1.57} & \textbf{1.94} & \textbf{2.42} \\

\midrule
ResLSTM & 1.51 & 1.07 & 1.19 & 1.46 & 1.90 & 2.44 & 1.10 & 1.20 & 1.39 & 1.70 & 2.18 & 1.08 & 1.26 & 1.51 & 1.87 & 2.24 & 1.06 & 1.20 & 1.46 & 1.80 & 2.29 \\
DenseNet & 1.48 & 1.05 & 1.15 & 1.39 & 1.83 & 2.32 & 1.06 & 1.14 & 1.32 & 1.56 & 2.02 & 1.09 & 1.26 & 1.51 & 1.82 & 2.23 & 1.05 & 1.15 & 1.43 & 1.78 & 2.23 \\
TCN-BC & 1.53 & 1.06 & 1.20 & 1.46 & 1.85 & 2.31 & 1.09 & 1.21 & 1.44 & 1.75 & 2.18 & 1.14 & 1.27 & 1.52 & 1.84 & 2.22 & 1.10 & 1.27 & 1.53 & 1.88 & 2.29 \\
TCN-BK & 1.51 & 1.09 & 1.24 & 1.53 & 1.98 & 2.46 & 1.11 & 1.24 & 1.52 & \textbf{1.86} & 2.22 & 1.17 & 1.38 & 1.65 & 1.92 & 2.29 & 1.14 & 1.29 & 1.54 & 1.88 & 2.30 \\


Prop. MB-TCN & 1.43 & \textbf{1.10} & \textbf{1.25} & \textbf{1.57} & \textbf{2.01} & \textbf{2.50} & \textbf{1.13} & \textbf{1.28} & \textbf{1.53} & 1.81 & \textbf{2.30} & \textbf{1.21} & \textbf{1.42} & \textbf{1.67} & \textbf{2.04} & \textbf{2.44} & \textbf{1.15} & \textbf{1.33} & \textbf{1.58} & \textbf{1.96} & \textbf{2.41} \\

\midrule
ResLSTM & 2.03 & 1.08 & 1.20 & 1.48 & 1.96 & 2.50 & 1.09 & 1.20 & 1.42 & 1.76 & 2.24 & 1.09 & 1.25 & 1.50 & 1.79 & 2.17 & 1.08 & 1.23 & 1.50 & 1.87 & 2.37 \\
DenseNet & 1.94 & 1.06 & 1.18 & 1.44 & 1.87 & 2.33 & 1.06 & 1.18 & 1.42 & 1.70 & 2.14 & 1.09 & 1.28 & 1.53 & 1.83 & 2.20 & 1.05 & 1.21 & 1.54 & 1.93 & 2.35 \\
TCN-BC & 2.03 & 1.06 & 1.19 & 1.42 & 1.86 & 2.31 & 1.07 & 1.19 & 1.41 & 1.72 & 2.20 & 1.11 & 1.27 & 1.50 & 1.82 & 2.22 & 1.07 & 1.23 & 1.47 & 1.80 & 2.20 \\
TCN-BK& 1.98 & 1.07 & 1.20 & 1.53 & 1.96 & 2.52 & 1.10 & 1.26 & 1.52 & 1.89 & \textbf{2.41} & 1.18 & 1.40 & 1.67 & 2.00 & 2.35 & 1.11 & 1.28 & 1.56 & 1.93 & 2.41 \\


Prop. MB-TCN & 1.66 & \textbf{1.11} & \textbf{1.27} & \textbf{1.58} & \textbf{2.05} & \textbf{2.54} & \textbf{1.14} & \textbf{1.28} & \textbf{1.52} & \textbf{1.91} & 2.33 & \textbf{1.21} & \textbf{1.43} & \textbf{1.72} & \textbf{2.13} & \textbf{2.51} & \textbf{1.18} & \textbf{1.34} & \textbf{1.61} & \textbf{1.97} & \textbf{2.42} \\
\midrule
LSTM-IRM  & 53.9 & 1.06 & 1.13 & 1.31 & 1.65 & 2.12 & 1.06 & 1.14 & 1.30 & 1.55 & 2.04 & 1.08 & 1.19 & 1.36 & 1.60 & 1.97 & 1.04 & 1.10 & 1.25 & 1.55 & 2.00 \\
BLSTM-IRM  & 39.0 & 1.07 & 1.17 & 1.40 & 1.81 & 2.24 & 1.08 & 1.17 & 1.36 & 1.65 & 2.07 & 1.09 & 1.23 & 1.43 & 1.73 & 2.12 & 1.05 & 1.14 & 1.34 & 1.66 & 2.07 \\
Prop. MB-TCN & 1.66 & \textbf{1.11} & \textbf{1.27} & \textbf{1.58} & \textbf{2.05} & \textbf{2.54} & \textbf{1.14} & \textbf{1.28} & \textbf{1.52} & \textbf{1.91} & \textbf{2.33} & \textbf{1.21} & \textbf{1.43} & \textbf{1.72} & \textbf{2.13} & \textbf{2.51} & \textbf{1.18} & \textbf{1.34} & \textbf{1.61} & \textbf{1.97} & \textbf{2.42} \\
        \bottomrule          
    \end{tabular}
    \label{taba}
\end{table*}

\begin{table*}[h!]
    \centering
    \scriptsize
  \def\arraystretch{1.35}
    \setlength{\tabcolsep}{3.6pt}
    \caption{
    Speech enhancement performance comparisons of different networks in terms of STOI (in \%) metric. The highest STOI score obtained at each condition and for each parameter size is highlighted with bold text.
    }
    
    \begin{tabular}{lc|lllll|lllll|lllll|lllll} 
        \toprule
        \multirow{3}{*}{\begin{tabular}[c]{@{}c@{}}\textbf{Network}\end{tabular}} &
        \multirow{3}{*}{\begin{tabular}[c]{@{}c@{}}\textbf{\# params.}\\$\pmb{\times 10^{6}}$\end{tabular}} &
        \multicolumn{20}{c}{\bf SNR level (dB)} \\
        \cline{3-22}
        & & \multicolumn{5}{c|}{\bf Voice babble} & \multicolumn{5}{c|}{\bf Street music} & \multicolumn{5}{c|}{\bf F16} & \multicolumn{5}{c}{\bf Factory} \\
        \cline{3-22}
        & & {\bf-5}  & {\bf0}   & {\bf5}   & {\bf10}  & {\bf15}        & {\bf-5}  & {\bf0}   & {\bf5}   & {\bf10}  & {\bf15}        & {\bf-5}  & {\bf0}   & {\bf5}   & {\bf10}  & {\bf15}  & {\bf-5}  & {\bf0}   & {\bf5}   & {\bf10}  & {\bf15} \\
        \hline
Noisy speech & - & 60.2 & 72.4 & 83.0 & 90.7 & 95.5 & 59.0 & 70.9 & 81.9 & 90.3 & 95.6 & 60.4 & 71.8 & 82.4 & 90.5 & 95.7 & 57.8 & 69.9 & 80.9 & 89.2 & 94.5\\ 
\midrule
ResLSTM & 1.02 & 61.0 & 75.8 & 86.8 & 93.4 & 96.7 
              & 63.0 & 77.1 & 86.7 & 92.8 & 96.3 
              & 66.1 & 78.4 & 87.1 & 92.9 & 96.6 
              & 62.1 & 77.4 & 87.0 & 92.7 & 96.2 \\
DenseNet & 0.97 & 58.3 & 73.1 & 86.0 & 93.0 & 96.5 
                & 59.1 & 74.0 & 84.7 & 92.0 & 96.2 
                & 65.3 & 77.8 & 86.7 & 92.4 & 95.9 
                & 56.8 & 73.0 & 85.1 & 91.7 & 95.9\\ 
TCN-BC & 1.03 & 60.8 & 75.5 & 87.2 & 93.4 & 96.6 
                  & 64.0 & 77.8 & 87.1 & 93.4 & 96.8 
                  & 67.2 & 79.4 & 87.9 & 93.4 & 96.9 
                  & 61.1 & 76.9 & 87.1 & 92.8 & 96.3 \\
TCN-BK & 1.05 & \textbf{62.6} & \textbf{78.1} & 88.4 & 94.2 & 97.0 
                  & 66.0 & 79.7 & 88.5 & 94.0 & 97.0 
                  & 68.4 & 80.8 & 88.7 & 93.9 & 97.1 
                  & \textbf{64.9} & 79.1 & 88.2 & 93.3 & 96.4 \\


Prop. MB-TCN & 1.05 & 62.1 & \textbf{78.1} & \textbf{88.6} & \textbf{94.3} & \textbf{97.1} 
                      & \textbf{66.8} & \textbf{80.0} & \textbf{89.2} & \textbf{94.3} & \textbf{97.2} 
                      & \textbf{69.2} & \textbf{82.2} & \textbf{89.7} & \textbf{94.5} & \textbf{97.4} 
                      & 64.7 & \textbf{80.0} & \textbf{88.3} & \textbf{93.4} & \textbf{96.5} \\

\midrule
ResLSTM & 1.51 & 59.6 & 74.6 & 86.3 & 93.0 & 96.6 & 62.9 & 76.9 & 86.8 & 92.8 & 96.4 & 64.0 & 77.4 & 87.4 & 93.1 & 96.3 & 61.0 & 76.6 & 86.7 & 92.6 & 96.2\\ 

DenseNet & 1.48 & 58.1 & 73.7 & 85.9 & 93.1 & 96.6 & 58.2 & 73.3 & 84.8 & 91.9 & 96.1 & 65.9 & 78.3 & 86.9 & 92.6 & 95.9 & 57.5 & 73.0 & 84.8 & 91.5 & 95.4\\ 

TCN-BC & 1.53 & 59.8 & 76.5 & 87.4 & 93.5 & 96.7 & 65.0 & 77.9 & 88.0 & 93.4 & 96.7 & 64.9 & 77.6 & 87.0 & 93.4 & 96.9 & 63.3 & 79.1 & 87.6 & 92.8 & 96.3\\ 

TCN-BK & 1.51 & 64.1 & 79.2 & 88.9 & 94.2 & 97.0 & 66.2 & 79.8 & 88.6 & 93.9 & 97.1 & 69.5 & 81.8 & 89.5 & 94.3 & 97.2 & 65.6 & 80.3 & 88.2 & 93.1 & 96.5\\ 


Prop. MB-TCN & 1.43 & \textbf{64.7} & \textbf{79.4} & \textbf{89.1} & \textbf{94.4} & \textbf{97.2} & \textbf{69.2} & \textbf{81.7} & \textbf{89.5} & \textbf{94.3} & \textbf{97.2} & \textbf{71.0} & \textbf{83.1} & \textbf{90.1} & \textbf{94.8} & \textbf{97.5} & \textbf{67.4} & \textbf{80.7} & \textbf{88.4} & \textbf{93.3} & \textbf{96.6} \\

\midrule
ResLSTM & 2.03 & 62.7 & 76.6 & 87.3 & 93.7 & 96.9 & 65.9 & 78.7 & 87.9  & 93.6 & 96.9 & 65.8 & 79.3 & 87.6 & 93.3 & 96.6 & 62.4 & 77.2 & 87.0 & 92.7 & 96.4\\ 

DenseNet & 1.94 & 57.8 & 73.4 & 85.8 & 92.3 & 95.8 & 60.4 & 75.2 & 86.8 & 93.1 & 96.6 & 63.4 & 77.8 & 87.1 & 92.9 & 96.0 & 58.4 & 74.4 & 86.4 & 92.6 & 96.2 \\

TCN-BC & 2.03 & 60.3 & 76.3 & 87.4 & 93.6 & 96.7 & 63.1 & 78.0 & 87.4 & 93.3 & 96.8 & 67.6 & 78.9 & 87.5 & 93.5 & 97.1 & 60.5 & 77.6 & 87.2 & 92.7 & 96.1 \\

TCN-BK & 1.98 & 59.0 & 74.1 & 87.3 & 93.8 & 97.1 & 65.3 & 80.6 & 89.1 & 94.1 & \textbf{97.1} & 68.7 & 81.5 & 89.3 & 94.3 & 97.1 & 64.3 & 79.9 & 88.5 & 93.2 & \textbf{96.6} \\


Prop. MB-TCN & 1.66 & \textbf{64.0} & \textbf{79.0} & \textbf{89.3} & \textbf{94.5} & \textbf{97.2} & \textbf{69.5} & \textbf{81.7} & \textbf{89.3} & \textbf{94.2} & \textbf{97.1} & \textbf{70.6} & \textbf{82.7} & \textbf{90.0} & \textbf{94.7} & \textbf{97.4} & \textbf{67.0} & \textbf{81.2} & \textbf{88.6} & \textbf{93.3} & \textbf{96.6} \\

\midrule
LSTM-IRM & 53.9 & 63.0 & 76.0 & 85.6 & 92.2 & 95.9 & 64.7 & 76.5 & 85.5 & 91.7 & 95.9 & 69.4 & 79.6 & 87.1 & 92.6 & 95.9 & 61.2 & 75.5 & 85.1 & 91.6 & 95.6 \\
BLSTM-IRM & 39.0 & \textbf{64.3} & 77.7 & 87.1 & 93.1 & 96.4 & 67.7 & 78.3 & 86.7 & 92.7 & 96.5 & 69.3 & 79.7 & 87.8 & 93.2 & 96.7 & 63.9 & 78.5 & 86.9 & 92.4 & 95.9 \\
Prop. MB-TCN & 1.66 & 64.0 & \textbf{79.0} & \textbf{89.3} & \textbf{94.5} & \textbf{97.2} & \textbf{69.5} & \textbf{81.7} & \textbf{89.3} & \textbf{94.2} & \textbf{97.1} & \textbf{70.6} & \textbf{82.7} & \textbf{90.0} & \textbf{94.7} & \textbf{97.4} & \textbf{67.0} & \textbf{81.2} & \textbf{88.6} & \textbf{93.3} & \textbf{96.6} \\
\bottomrule          
    \end{tabular}
    \label{tabb}
\end{table*}

%% file: composite.tex
\begin{table*}[t!]
    \centering
    \footnotesize	
    \def\arraystretch{1.2}
    \setlength{\tabcolsep}{3.5pt}
    \caption{Comparison with other state-of-the-art speech enhancement models on the second publicly available dataset. Higher score (CSIG, CBAK, COVL, PESQ, and STOI) indicates better performance and the highest scores obtained for each evaluation measure are highlighted with bold text. For comparison of parameter efficiency, we present the number of trainable parameters in different models.}
    
    \begin{tabular}{@{}lcclllll@{}}
    \toprule
    \textbf{Methods} & \textbf{\# Parameters (M)} &  \textbf{Types} & \textbf{CSIG} & \textbf{CBAK} & \textbf{COVL} & \textbf{PESQ} & \textbf{STOI} (in \%)\\
    \midrule
    \midrule
Noisy Input & - &  - & 3.35  & 2.44 & 2.63 & 1.97 & 92 (91.5)\\
Wiener (\cite{wiener1996}, Scalart et al. 1996) & - &  T-F domain & 3.23 & 2.68 & 2.67 & 2.22 & - \\ 
SEGAN (\cite{SEGAN2017}, Pascual et al. 2017) & 43.2 M (25.8 M) & Time-domain & 3.48 & 2.94 & 2.80 & 2.16 & 93 \\ 
Wavenet (\cite{wavenet2018}, Rethage et al. 2018) & 6.34 M &  Time-domain & 3.62 & 3.23 & 2.98 & - & -\\ 
Wave-U-Net (\cite{waveunet2018}, Macartney et al. 2018) & 10.2 M & Time-domain & 3.52 & 3.24 & 2.96 & 2.40 & -\\
Deep Feature Loss(\cite{deeploss2018}, Germain et al. 2018) & 0.64 M & Time-domain & 3.86 & 3.33 & 3.22 & - & -\\
MMSE-GAN (\cite{MMSE-GAN2018}, Soni et al. 2018)  & 0.79 M (0.56 M) &  T-F domain & 3.80 & 3.12 & 3.14 & 2.53 & 93\\ 
Metric-GAN (\cite{metricGAN2019}, Fu et al. 2019) & 1.89 M (0.35 M) &  T-F domain & 3.99 & 3.18 & 3.42 & 2.86 & -\\
MDPhD (\cite{MDPhD2018}, Kim et al. 2018) & 6 M & Hybrid & 3.85 & 3.39 & 3.27 & 2.70 & -\\
\midrule
\midrule
Proposed MB-TCN (DeepXi - SRWF) & \multirow{3}{*}{1.66 M}  & T-F domain & 4.20 & 3.32 & 3.55 & 2.87 & {\bf 94} ({\bf 93.70}) \\
Proposed MB-TCN (DeepXi - MMSE-STSA) &  & T-F domain & {\bf 4.21} & 3.36 & 3.57 & 2.91 & {\bf 94} (\bf 93.70)\\
Proposed MB-TCN (DeepXi - MMSE-LSA) &  & T-F domain & {\bf 4.21} & {\bf 3.41} & {\bf 3.59} & {\bf 2.94} & {94} (93.64) \\
\bottomrule

\end{tabular}
\label{tab:composite}
\end{table*}